\documentclass[aps, prl, showpacs, superscriptaddress ,twocolumn ]{revtex4}
\usepackage{amsmath}
\usepackage{graphicx}
\usepackage{sidecap}
\usepackage{hyperref}

\newcommand{\LSCO}{La$_{2-x}$Sr$_x$CuO$_4$}

\begin{document}

\title{Evolution from a nodeless gap to $d_{x^2-y^2}$ form in underdoped  La$_{2-x}$Sr$_x$CuO$_4$}

\author{E. Razzoli}
\affiliation{Swiss Light Source, Paul Scherrer Institute, CH-5232 Villigen PSI, Switzerland}

\author{G. Drachuck}
\affiliation{Department of Physics, Technion, Haifa 32000, Israel}

\author{A. Keren}
\affiliation{Department of Physics, Technion, Haifa 32000, Israel}

\author{M. Radovic}
\affiliation{Swiss Light Source, Paul Scherrer Institute, CH-5232 Villigen PSI, Switzerland}
\affiliation{Institut de la Matiere Complexe, EPF Lausanne, CH-1015, Lausanne, Switzerland}

\author{N. C. Plumb}
\affiliation{Swiss Light Source, Paul Scherrer Institute, CH-5232 Villigen PSI, Switzerland}

\author{J. Chang}
\affiliation{Swiss Light Source, Paul Scherrer Institute, CH-5232 Villigen PSI, Switzerland}
\affiliation{Institut de la Matiere Complexe, EPF Lausanne, CH-1015, Lausanne, Switzerland}

\author{J. Mesot}
\affiliation{Swiss Light Source, Paul Scherrer Institute, CH-5232 Villigen PSI, Switzerland}
\affiliation{Institut de la Matiere Complexe, EPF Lausanne, CH-1015, Lausanne, Switzerland}

\author{M. Shi}
\affiliation{Swiss Light Source, Paul Scherrer Institute, CH-5232 Villigen PSI, Switzerland}

\begin{abstract}

Using angle-resolved photoemission (ARPES), it is revealed that the
low-energy electronic excitation spectra of highly underdoped
superconducting and non-superconducting La$_{2-x}$Sr$_x$CuO$_4$
cuprates are gapped along the entire underlying Fermi surface at low
temperatures. We show how the gap function evolves to a
$d_{x^2-y^2}$ form as increasing temperature or doping, consistent
with the vast majority of ARPES studies of cuprates. Our results
provide essential information for uncovering the symmetry of the
 order parameter(s) in strongly underdoped cuprates,
which is a prerequisite for understanding the pairing mechanism and
how superconductivity emerges from a Mott insulator.

\end{abstract}

\date{\today}
\maketitle

In the Bardeen-Cooper-Schriefer (BCS) theory of superconductivity,
the symmetry of the superconducting gap reflects the order parameter
of the superfluid state and is directly tied to the symmetry of the
interactions driving the formation of Cooper pairs. Similarly, other
ordered phases, such as charge- or spin-density wave states, can
induce gaps whose symmetries are connected to the underlying
order parameters. Thus in high-temperature superconductors, where
superconductivity is found in close proximity to magnetic and charge
order, the symmetry of the gap function is of critical theoretical
importance. It is now widely accepted that the superconducting gap
in moderately hole-doped high-temperature superconducting copper
oxides (cuprates) exhibits a node located along the diagonal
(0,0)-($\pi$,$\pi$) line of the Brillouin zone (BZ)
\cite{Damascelli2003, Chatterjee2010}, consistent with an overall
gap function of pure $d_{x^2-y^2}$ symmetry \cite{Tsuei2000}. A key
issue is whether $d_{x^2-y^2}$ is the only form of the gap function
for all cuprates over the full range of dopings.
Using angle-resolved photoemission (ARPES) we reveal that the
low-energy electronic excitation spectra of highly underdoped
superconducting and non-superconducting La$_{2-x}$Sr$_x$CuO$_4$
(LSCO) are gapped along the entire underlying Fermi surface (FS) at
low temperatures. On the zone diagonal, gapless excitations appear
as the temperature and/or doping is increased, and the gap function 
evolves to a $d_{x^2-y^2}$ form.

High quality single crystals of superconducting and
non-superconducting \LSCO~were grown using the traveling solvent
floating zone method. ARPES experiments were carried out at the
Surface and Interface Spectroscopy beamline at the Swiss Light
Source (SLS). Circularly polarized light with $h\nu = 55$ eV was
used. The spectra were recorded with a Scienta R4000 electron
analyzer. The energy resolutions was about 14 - 17 meV.

\begin{SCfigure*}
\includegraphics[width=0.65\textwidth]{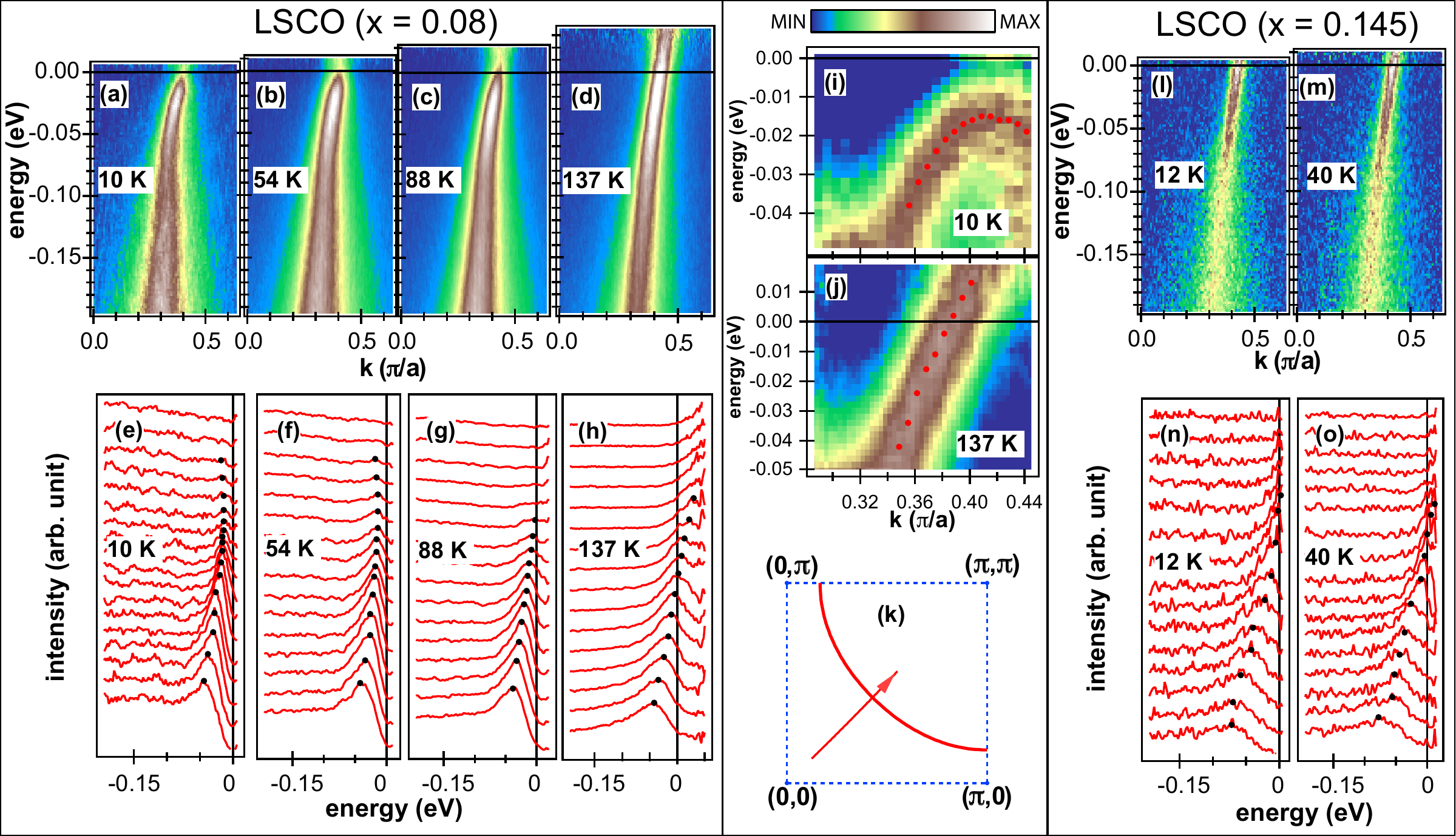}
\caption{(Color online)  ARPES spectra for LSCO with $x = 0.08$
($T_c = 20$ K) and $x = 0.145$ ($T_c = 33$ K). (a)-(d) Intensities
along the zone diagonal [red arrow in (k)] at $T = 10$ K, 54 K, 88 K and
137 K for $x = 0.08$. The spectra were obtained by the DFD method 
\cite{Yang2008}. (e)-(h) EDCs from (a)-(d) in the vicinity of $k_F$. 
(i)-(j) Images in the vicinity of $k_F$ at 10 K and 137 K. Each EDC is
normalized to the intensity at the peak position (red circles). (k) The first
quadrant of the BZ. The red curve is the FS of LSCO $x = 0.08$ obtained
from tight-binding (TB) fits to the experimental data. The red arrow
indicates the cut along which the ARPES data were taken. (l)-(m) the
same as (a)-(d) but for LSCO ($x = 0.145$) at $T = 12$ K and 40 K.
(n)-(o) EDCs from (l)-(m) close to $k_F$. }\label{T_dep}
\end{SCfigure*}

In Fig. \ref{T_dep}(a)-(d) we show ARPES spectra below and above the
superconducting transition temperature ($T_c$) for highly underdoped
superconducting LSCO~($ x = 0.08$, $T_c =20$ K) along the diagonal 
line of the BZ. The spectra were obtained by deconvoluting the raw 
ARPES data to remove the broadening due to the finite instrumental
resolution and then dividing the deconvoluted spectra by a Fermi
distribution function [deconvolution - Fermi function division (DFD)
method] \cite{Yang2008}. Relative to $E_F$, a gap is clearly
observed both below $T_c$ (10 K) and above $T_c$ (54 K). The gap
closes above  $\sim 88$ K. To reveal the details of the gap, we
trace the dispersion in the vicinity of $E_F$. In Fig.
\ref{T_dep}(e)-(h) we plot energy distribution curves (EDCs) from
Fig. \ref{T_dep}(a)-(d) along the zone diagonal cut. At low
temperatures (10 K and 54 K), moving from (0,0) to ($\pi$,$\pi$),
the peak position of the EDCs approaches $E_F$, but before reaching
$E_F$ it recedes to higher binding energies (Fig.
\ref{T_dep}(e)-(f)). The spectral peak reaches $E_F$ at 88 K (Fig.
\ref{T_dep}(g)) and crosses $E_F$ at higher temperatures [Fig.
\ref{T_dep}(h)], signaling the closure of the gap. In Fig. \ref{T_dep} 
(i)-(j) the dispersion bending back below $E_F$ at 10 K and crossing 
$E_F$ at 137 K are better visualized by normalizing each EDC to its
peak intensity.

In contrast to the observation of an energy gap along
(0,0)-($\pi$,$\pi$) at low temperatures for LSCO ($x = 0.08$), we
did not find such a gap in the spectra of the optimally doped LSCO
($x = 0.145$, $T_c = 33$ K) sample. The spectral peak in this higher
hole-doped sample continuously moves to higher energies and crosses
$E_F$, even at temperatures down to 12 K [Fig. \ref{T_dep}(l)-(o)].

%

\begin{figure}
\includegraphics[width=0.49\textwidth]{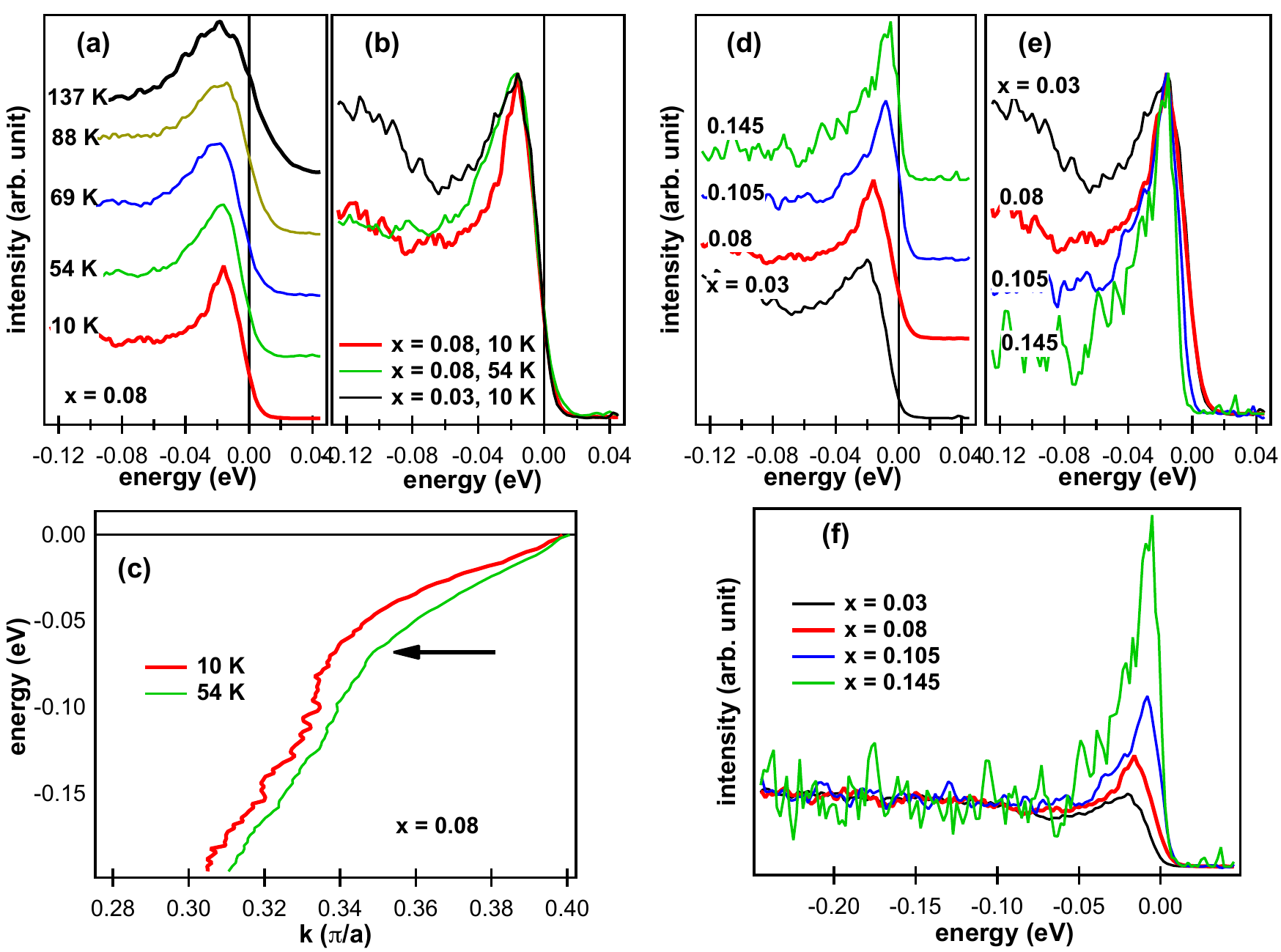}
\caption{(Color online) EDCs at $k_F$ on the zone diagonal and band
dispersions along the zone diagonal for LSCO. The EDCs were obtained
by deconvoluting the raw ARPES data to remove the broadening due to
the finite instrumental resolution \cite{Yang2008}. (a) EDCs at
$k_F$ as a function of temperature for $x = 0.08$. Curves are offset
vertically for clarity. (b) EDCs at $k_F$ for $x = 0.03$ at 10 K,
and for $x = 0.08$ at 10 K and 54 K. Curves are offset horizontally
to align the peak position to that of the EDC for $x = 0.08$ at 10
K. (c) Dispersions derived from the peak positions of the momentum
distribution curves for $x = 0.08$ at 10 K ($< T_c$) and 54 K ($>
T_c$). The kink is indicated by the black arrow. (d)-(e) EDCs at 10
K as a function of doping ($x$). Curves in (d) are offset vertically
for clarity and in (e) are offset horizontally to align the peak
position to that of the EDC for $x = 0.08$ at 10 K. (f) The same
EDCs as in (d), but the spectra are normalized to intensities at
high binding energy.}\label{width}
\end{figure}

A general observation in cuprates is that when $T_c$ is crossed by
decreasing the temperature, EDC peaks near the Fermi momentum
($k_F$) along an off-nodal cut sharpen, an energy gap opens at
$k_F$, and the renormalization of the band dispersion becomes more
pronounced. This behavior was also observed in our ARPES spectra for
LSCO ($x = 0.08$) along the diagonal cut [red arrow in Fig.
\ref{T_dep} (k)]. In Fig. \ref{width}(a)-(b) we plot the EDCs at the
$k_F$ on the zone diagonal line as a function of temperature. The
EDC peak width at 10 K ($< T_c$) is considerably smaller than when
the spectrum is measured above $T_c$ ($T = 54$ K) [Fig.
\ref{width}(b)]. At low temperature (10 K) the peak width of the
superconducting sample LSCO ($x = 0.08$) is also smaller than that
of the non-superconducting sample ($x = 0.03$) [Fig.
\ref{width}(b)], which indicates the change in the peak width is
associated with the superconducting transition. The anomalous ``kink''
in the dispersion along the nodal direction, which is well-known
from previous ARPES studies \cite{Kaminski2001, Lanzara2001, Gromko2003},
becomes more pronounced below $T_c$ [Fig. \ref{width}(c)], showing
that the renormalization is enhanced in the superconducting state.
To gain more insight about the differences between the spectra in
the superconducting and non-superconducting phases, Fig.
\ref{width}(d)-(f) shows the EDCs as a function of doping at 10 K. The EDC peak
positions of the $x = 0.03$ and 0.08 samples occur at higher binding energy
than in the $x = 0.105$ and 0.145 samples [Fig. \ref{width}(d)], in which a simple
$d_{x^2-y^2}$ superconducting gap was observed \cite{Shi2008,
Shi2009}. An energy gap of $\sim 20$ meV along the
(0,0)-($\pi$,$\pi$) is also seen in the non-superconducting sample ($x
= 0.03$) at 10 K. For the superconducting samples, after
aligning the peaks to the same position, the falling edges of the
EDC peaks are almost identical at low binding energies [Fig.
\ref{width}(e)], which demonstrates that the coherent peaks have the
same width in the superconducting state, and this width is smaller
than that of the non-superconducting sample. We observe that underdoping increases 
the transfer of spectral weight from the coherent peak to high binding
energies [Fig. \ref{width}(f)].

\begin{figure}
\includegraphics[width=0.49\textwidth]{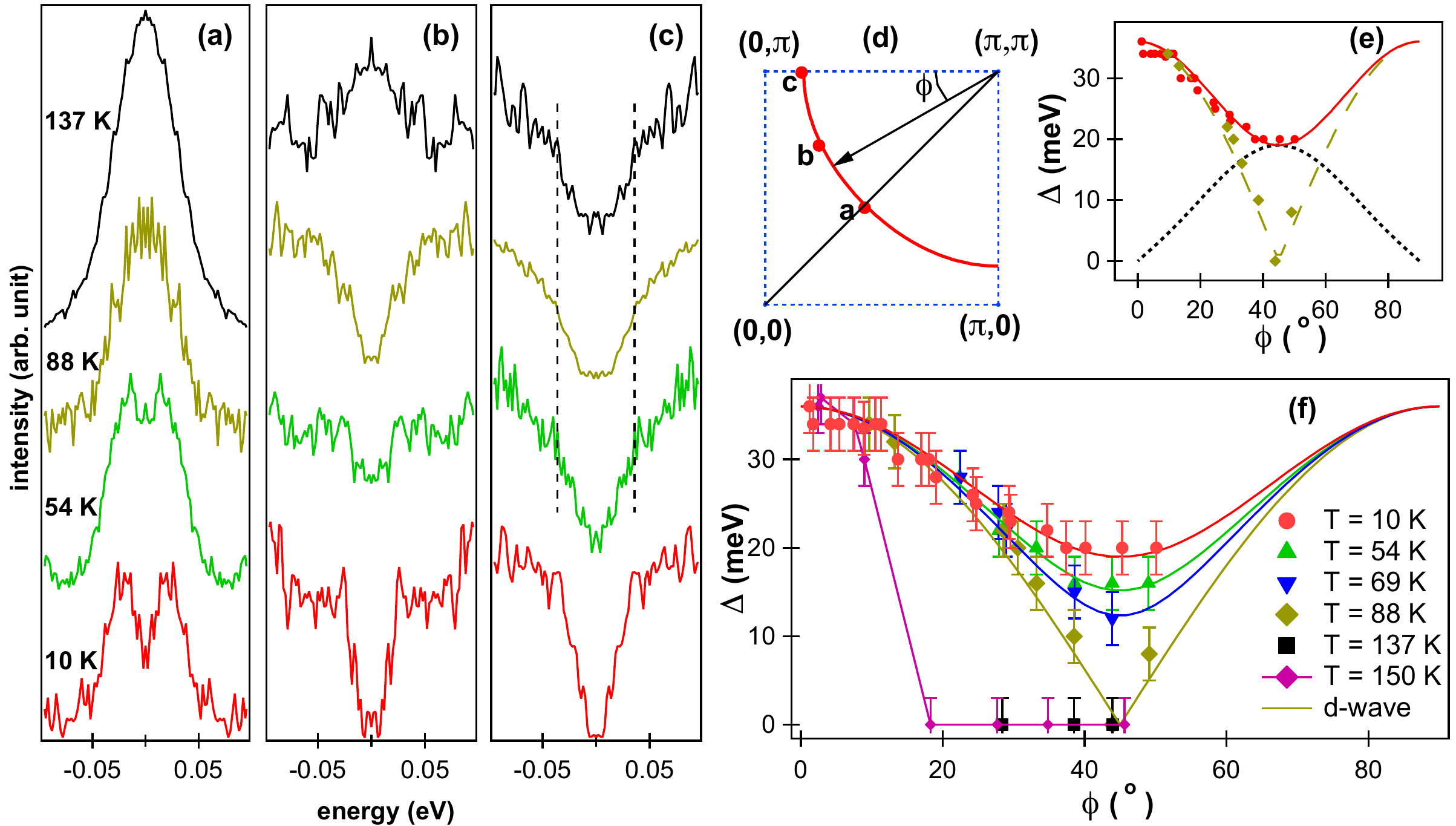}
\caption{(Color online) ARPES spectra for LSCO ($x = 0.08$). (a) -
(c) Symmetrized EDCs as a function of temperature at $k_F$ (a), (b)
and (c) shown in (d). (d) The first quadrant of the BZ. The red
curve is the FS. The closed circles indicate the $k_F$ at which the
symmetrized EDCs are shown in (a)-(c). (e) Experimentally determined
energy gap as a function of the FS angle $\phi$ indicated in (d)
at 10 K (circles) and 89 K (diamonds). The dash line is the
$d_{x^2-y^2}$ form in Eq.~(\ref{eq1}) with
$\Delta_{d_{x^2-y^2}}^0 = 38.52$ meV. The dotted line is the
$d_{xy}$ form in Eq.~(\ref{eq2}) with $\Delta_{d_{xy}}^0 = 20$
meV. The solid line is $|\Delta_{d_{x^2-y^2}} + i\Delta_{d_{xy}}|$.
(f) Energy gap as a function of $\phi$. The symbols are the
experimentally determined energy gaps at various temperatures.
The solid lines are $|\Delta_{d_{x^2-y^2}} +
i\Delta_{d_{xy}}|$ with $\Delta_{d_{x^2-y^2}}^0 = 38.52$ meV and
$\Delta_{d_{xy}}^0 = 20$ meV (red), 16.6 meV (green), 13.5 meV
(blue), and 0 meV (yellow-green). }\label{Gap}
\end{figure}

The momentum dependence of the gap as a function of temperature is
shown in Fig. \ref{Gap}. The measured raw EDCs at $k_F$ were
symmetrized to remove the effects of the Fermi
function~\cite{Norman1998}. In the superconducting state, the
extracted gap sizes were defined as half the peak-to-peak separation
of the symmetrized EDCs. Above $T_c$, for those spectra
having no coherent peak, the gap was defined as half the distance
between the two locations where the slope has the largest change, as
indicated by vertical lines in Fig. \ref{Gap}(c). For LSCO ($x =
0.08$), the energy gap is highly anisotropic [Fig. \ref{Gap}(f)]. It
has a maximal value at the zone boundary ($\phi = 0^{\circ}$) and
decreases monotonically along the FS to a minimum at the zone
diagonal ($\phi = 45^{\circ}$). However, at low temperatures, the
gap ($\Delta$) function strongly deviates from a pure
$d_{x^2-y^2}$ form,
\begin{equation}
\Delta_{d_{x^2-y^2}} (\boldsymbol{k}) = {\Delta_{d_{x^2-y^2}}^0} [\cos( k_xa) -\cos( k_ya)]/2,  \label{eq1}
\end{equation}
where $a$ is the lattice constant. Below $T_c$ at the $k_F$ on zone diagonal, a finite gap (hereafter we use the term ``diagonal gap'' for simplicity) is observed, which has an amplitude of $\sim 20$ meV. As the temperature is increased, the diagonal gap monotonically decreases. It disappears at $\sim 88$ K, at which point the gap function is very close to a pure $d_{x^2-y^2}$ form [Fig. \ref{Gap}(f)].  At higher temperatures, gapless excitations appear on a portion of the FS centered at the zone diagonal (Fermi arc) and the arc length increases with temperature, a phenomenon that has been observed in early ARPES work of underdoped Bi$_2$Sr$_2$CaCu$_2$O$_{8+\delta}$(Bi-2212) \cite{Kanigel2006} and other cuprates. In contrast to the temperature dependence of the diagonal gap, the gap size near the zone boundary is insensitive to temperature in the range of 10 K - 150 K; it always has a value of $\sim 36$ meV [Fig. \ref{Gap}(f)].
%
%
%
%
%

\begin{SCfigure*}
\includegraphics[width=0.65\textwidth]{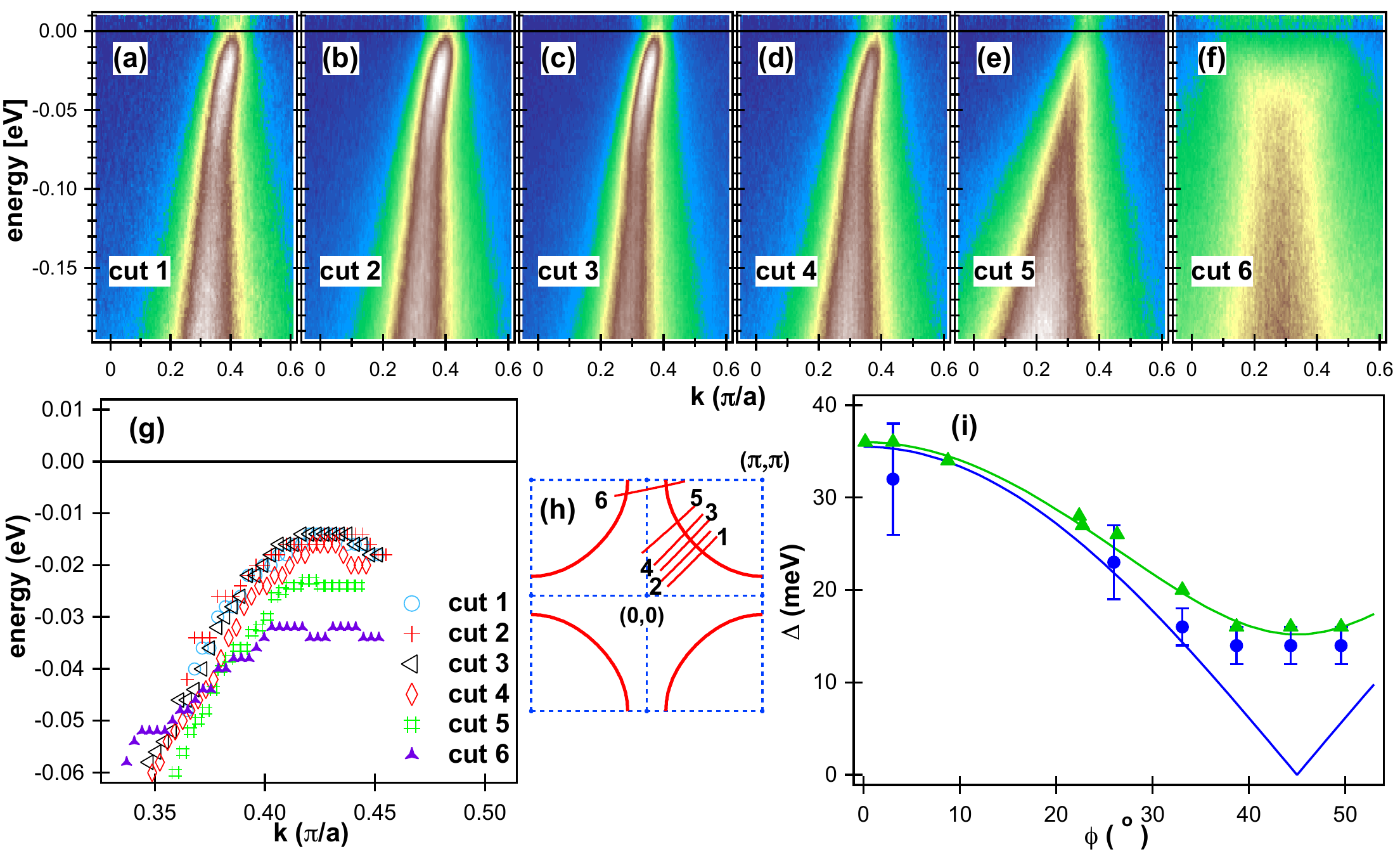}
\caption{(Color online) ARPES spectra of  LSCO ($x = 0.08$) measured
at 54 K ($T_c=20$ K). (a)-(f) Intensities along 6 cuts indicated in
(h). The spectra were obtained by the DFD method. (g) Dispersions obtained by tracing the EDC peaks
of (a)-(f) in the vicinity of $k_F$. For clarity the dispersions
along cuts 2-6 were offset horizontally. (h) The FS (red line)
obtained from TB fits to the experimental data. Lines 1-6 indicate
the cuts along which the data in (a)-(f) were taken. (i) Energy gap
as a function of FS angle ($\phi$). The closed circles are the gap
extracted from the difference between $E_F$ and the maximal energy
that the back-bending dispersions reach. The triangles are the gap
obtained from symmetrized EDCs [Fig.~\ref{Gap}(f)]. The solid lines
are the $|\Delta_{d_{x^2-y^2}}|$ (blue) and $|\Delta_{d_{x^2-y^2}} +
$i$\Delta_{d_{xy}}|$ (green) as in Fig.
\ref{Gap}(f).}\label{Gap_dec}
\end{SCfigure*}

The gap function obtained from the symmetrization method, in which
particle-hole symmetry is assumed, is confirmed with an independent
procedure. We have collected high-statistics ARPES spectra [Fig.
\ref{Gap_dec}(a)-(f)] along some selected cuts [Fig.
\ref{Gap_dec}(h)] and then applied the DFD method to trace the dispersion in the vicinity
of $E_F$. The energy gap was determined from the difference between
$E_F$ and the maximal energy that a back-bending dispersion reaches
[Fig.~\ref{Gap_dec} (g)]. Fig.~\ref{Gap_dec}(i) shows the results of
such an analysis applied to the data taken at 54 K. Within the error
bars, the obtained gap function is consistent with that derived from
the symmetrization method.

%
%
%

To explain our results, one natural line of thinking is to assume
that both the superconducting gap and the pseudogap have
$d_{x^2-y^2}$ form,  and the combination of disorder with
long-range Coulomb interactions depresses the density of
single-particle excitations at $E_F$, forming a so-called Coulomb gap
\cite{Shen2004, Pan2009}. A counter argument to this picture is that
the scattering due to the disorder should also reduce the lifetime
of quasiparticles, which would be manifested as severe broadening in
the ARPES lineshapes. Such broadening is not seen in the data, where 
instead the widths of the EDCs are the same in the superconducting
state for samples with different dopings. Moreover, these widths are
smaller than those at $T > T_c$ or in the non-superconducting sample
at low temperature (Fig.~\ref{width}).
One could instead consider a scenario in which an ordered phase
(e.g.,  a spin-density wave \cite{Doiron2012, Sachdev2010}) or its
flutuations above $T_c$ opens an energy gap along the entire FS in
lightly doped LSCO ($x < 0.105$). In such a picture, this
phase is different from the superconducting
instability, and below $T_c$ it coexists with the superconducting
phase, which has a $d_{x^2-y^2}$ order parameter. The end form
or the combination of these two order parameters would produce an
energy gap on the entire FS. Recent Monte Carlo simulations have
shown that a strong fluctuating competing order could open a
diagonal gap in the superconducting phase, and a $d_{x^2-y^2}$
gap function is restored with reduced amplitudes of the fluctuating
competing order~\cite{Andersen2012}. This scenario would explain why the diagonal gap is open in LSCO $x = 0.08$ and 0.03 at low temperatures and is closed for $x \geq  0.1$, since it has been suggested that disordered magnetism is present  at low doping and disappears at optimal doping \cite{Andersen2007, Andersen2010}. The open question is how to
account for the similar momentum dependence of the gap above and below $T_c$ in
LSCO ($x = 0.08$), as well as the fact that the amplitude of the gap at the anti-node
is relatively insensitive to temperature.
An alternative explanation is that in LSCO~ there is a critical
doping point below which the superconducting gap function changes
from $d_{x^2-y^2}$ to another form. The pseudogap at $T  > T_c$
is a precursor to superconductivity \cite{Emery1995, Randeria1992,
Gomes2007, Li2010}, which has similar momentum dependence and
amplitude as the superconducting gap. To study the symmetry of the
gap function in the superconducting state of the highly underdoped
LSCO ($x = 0.08$) sample, we have compared the obtained gap function
at 10 K [Fig.~\ref{Gap}(e)-(f)] with those of all allowed
spin-singlet even-parity pair states in cuprates
\cite{Annett1990,Tsuei2000}. The only order parameter basis function
that is nodeless in its pure form is the $s$-wave state. However,
although this order parameter is nodeless, its momentum dependence
is not compatible with the observed momentum dependence of the
energy gap. A combination of two real subcomponents of different
superconducting order parameters, such as a mixed $s + d_{x^2-y^2}$
gap \cite{Khasanov2007}, will only shift the node(s) along the FS,
which cannot produce a nodeless gap structure. By considering all
the possible nodeless mixed gap function we find that the
combination of $d_{x^2-y^2} +id_{xy}$ gaps is the only form that can
reproduce the momentum dependence of the measured gap function
quantitatively. In Fig.~\ref{Gap}(e) we plot $|\Delta_{d_{x^2-y^2}}
(\boldsymbol{k})|$ and $|\Delta_{d_{xy}}(\boldsymbol{k})|$, as well
as the modulus of their sum $\Delta(\boldsymbol{k}) =
|\Delta_{d_{x^2-y^2}} (\boldsymbol{k}) +
i\Delta_{d_{xy}}(\boldsymbol{k})|$, as a function of FS angle
($\phi$), where
\begin{equation}
\Delta_{d_{xy}}(\boldsymbol{k})=\Delta_{d_{xy}}^0 [\sin(k_xa)\sin(k_ya)]. \label{eq2}
\end{equation}
We take $\Delta_{d_{x^2-y^2}}^0 = 1.07 \Delta_{antinode}$ and
$\Delta_{d_{xy}}^0 = \Delta_{diag}$, where $ \Delta_{antinode}$ and
$\Delta_{diag}$ are the measured gap sizes at the zone boundary and
the zone diagonal, respectively. The pre-factor of 1.07 comes from
$2/[\cos(k_{Fx}a) - \cos(k_{Fy}a)]$ where $(k_{Fx},k_{Fy})$ is the
intersection of the FS and the zone boundary. Remarkably, excellent
agreement is found between the experimental data and the mixed
$d_{x^2-y^2} + id_{xy}$ gap function along the entire FS  for all
the temperatures.  To see this, we simply hold
$\Delta_{d_{x^2-y^2}}^0$  constant and use the measured gap sizes on the zone
diagonal $(\Delta_{diag})$ for the amplitude of the $d_{xy}$
contribution to the mixed  $d_{x^2-y^2} + id_{xy}$ gap at different
temperatures. The results are shown in comparison to the data in
Fig.~\ref{Gap}(f). Interestingly, the mixed  $d_{x^2-y^2} + id_{xy}$
pair state implies that the time-reversal symmetry is broken in the
system \cite{Tsuei2000}, which has been embodied in early theories
of quantum phase transitions \cite{Vojta2000, Varma2006}.

To summarize, our main experimental findings are: 1) for highly
underdoped LSCO ($x = 0.08$), in the superconducting state the
electronic excitations are gaped along the entire underlying FS; 2)
the diagonal gap persists above $T_c$ while increasing
the temperature and/or reducing the hole-concentration; 3) whereas the gap
size on zone boundary remains constant up to 150 K, the diagonal gap
is temperature-dependent; 4) when the diagonal gap decreases to
zero, it leaves behind a pure $d_{x^2-y^2}$ energy gap, at
which point a Fermi arc emerges and its length increases with
temperature. Our observations in highly underdoped LSCO could be
explained either by a strong fluctuating competing order which is
different from the superconducting order parameter \cite{Andersen2012}, or
by a mixed $d_{x^2-y^2} + id_{xy}$ gap function occurring when the doping 
is below a quantum critical point \cite{Vojta2000}. The entirely gapped Fermi
surface in highly underdoped LSCO is a profound departure from the 
$d_{x^2-y^2}$ form observed at moderate to high doping. Further 
investigations of the origins of this nodeless gap function and its evolution
to pure $d_{x^2-y^2}$ form promise to shed light on how superconductivity
emerges from the Mott insulating state in high-$T_c$ cuprates.

We thank M.R. Norman, T.M. Rice, M. Sigrist and C. Bernhard for
useful discussions. This work was supported by the Swiss National
Science Foundation (through MaNEP, grant No. 200020-105151) and by
the Israeli Science Foundation. We thank the beamline staff of X09LA
at the SLS for their excellent support.



\begin{thebibliography}{1}


\bibitem{Damascelli2003}  A. Damascelli,  Z. Hussain, Z.-X. Shen, Rev. Mod. Phys. \textbf{75}, 473 (2003).
\bibitem{Chatterjee2010} U. Chatterjee \textit{et al.}, Nature Phys. \textbf{6}, 99 (2010).
\bibitem{Tsuei2000} C. C. Tsuei, and J. R Kirtley, Rev. Mod. Phys. \textbf{72}, 969 (2000).
\bibitem{Yang2008}  H. B. Yang \textit{et al.}, Nature (London) \textbf{456}, 77 (2008).
\bibitem{Lanzara2001} A. Lanzara \textit{et al.}, Nature (London) \textbf{412}, 510--514 (2001).
\bibitem{Kaminski2001}  A. Kaminski \textit{et al.}, Phys. Rev. Lett. \textbf{86}, 1070 (2001).
\bibitem{Gromko2003} A. D. Gromko \textit{et al.}, Phys. Rev. B \textbf{68}, 174520 (2003).
\bibitem{Shi2008} M. Shi \textit{et al.},  Phys. Rev. Lett. \textbf{101}, 047002 (2008).
\bibitem{Shi2009} M. Shi  \textit{et al.}, Europhys. Lett. \textbf{88}, 27008 (2009).
\bibitem{Norman1998} M. R. Norman  \textit{et al.}, Nature (London) \textbf{392}, 157 (1998).
\bibitem{Kanigel2006} A. Kanigel  \textit{et al.},  Nature Phys. \textbf{2}, 447 (2006).
\bibitem{Shen2004} K. M. Shen \textit{et al.},  Phys. Rev. B \textbf{69}, 054503 (2004).
\bibitem{Pan2009} Z.-H. Pan \textit{et al.}, Phys. Rev. B \textbf{79}, 092507 (2009).
\bibitem{Doiron2012}  N. Doiron-Leyraud and L. Taillefer, preprint arXiv: 1204.0490v1 (2012).
\bibitem{Sachdev2010} S. Sachdev,  Physica C \textbf{470}, S4 (2010).
\bibitem{Andersen2012} W. A. Atkinson, D. J. Bazak and B. M. Andersen, preprint arXiv:1206.2313v1 (2012).
\bibitem{Andersen2007} B. M. Andersen, P. J. Hirschfeld, A. P. Kampf and M. Schmid, Phys. Rev. Lett. \textbf{99}, 147002 (2007).
\bibitem{Andersen2010} B. M. Andersen, S. Graser and P. J. Hirschfeld, Phys. Rev. Lett. \textbf{105}, 147002 (2010).
\bibitem{Emery1995} V. J. Emery and S. A. Kivelson, Nature (London) \textbf{374}, 434 (1995).
\bibitem{Randeria1992}  M. Randeria, N. Trivedi, A. Moreo, and R. T. Scalettar,  Phys. Rev. Lett. \textbf{69}, 2001 (1992).
\bibitem{Gomes2007} K. K. Gomes  \textit{et al.},  Nature (London) \textbf{447}, 569 (2007).
\bibitem{Li2010} L. Li  \textit{et al.},  Phys. Rev. B \textbf{81}, 054510 (2010).
\bibitem{Annett1990}  J.F. Annet, Adv. Phys. \textbf{39}, 83 (1990).
\bibitem{Khasanov2007}  R. Khasanov \textit{et al.}, Phys. Rev. Lett. \textbf{99}, 237601 (2007).
\bibitem{Vojta2000}  M. Vojta,  Y. Zhang, and S. Sachdev, Phys. Rev. Lett. \textbf{85}, 4940 (2000).
\bibitem{Varma2006} C. M. Varma,  Phys. Rev. B \textbf{73}, 155113 (2006).

\end{thebibliography}
\end{document}